\documentclass[runningheads]{llncs}

\usepackage{graphicx}
\usepackage{color}
\usepackage[colorlinks=false]{hyperref}
\usepackage{soul}
\usepackage{amssymb}
\usepackage{mwe}
\usepackage{float}
\usepackage{array}
\usepackage[british]{babel}
\usepackage{multirow}
\usepackage{makecell}
\usepackage{caption}
\usepackage{subcaption}
\newcolumntype{L}[1]{>{\raggedright\let\newline\\\arraybackslash\hspace{0pt}}m{#1}}
\newcolumntype{C}[1]{>{\centering\let\newline\\\arraybackslash\hspace{0pt}}m{#1}}
\newcolumntype{R}[1]{>{\raggedleft\let\newline\\\arraybackslash\hspace{0pt}}m{#1}}

\begin{document}

\title{Uncovering the Semantics\\of Wikipedia Categories}
\author{
	Nicolas Heist\orcidID{0000-0002-4354-9138} \and
	Heiko Paulheim\orcidID{0000-0003-4386-8195}
}
\authorrunning{N. Heist, H. Paulheim}

\institute{
	Data and Web Science Group, University of Mannheim, Germany
	\email{\{nico,heiko\}@informatik.uni-mannheim.de}
}

\maketitle

\begin{abstract}
The Wikipedia category graph serves as the taxonomic backbone for large-scale knowledge graphs like YAGO or Probase, and has been used extensively for tasks like entity disambiguation or semantic similarity estimation. Wikipedia's categories are a rich source of taxonomic as well as non-taxonomic information. The category \emph{German science fiction writers}, for example, encodes the type of its resources (\emph{Writer}), as well as their nationality (\emph{German}) and genre (\emph{Science Fiction}). Several approaches in the literature make use of fractions of this encoded information without exploiting its full potential.
In this paper, we introduce an approach for the discovery of category axioms that uses information from the category network, category instances, and their lexicalisations. With DBpedia as background knowledge, we discover 703k axioms covering 502k of Wikipedia's categories and populate the DBpedia knowledge graph with additional 4.4M relation assertions and 3.3M type assertions at more than 87\% and 90\% precision, respectively.

\end{abstract}

\section{Introduction} \label{introduction}
Two of the most prominent public knowledge graphs, DBpedia \cite{lehmann2015dbpedia} and YAGO \cite{mahdisoltani2013yago3}, build rich taxonomies using Wikipedia's infoboxes and category graph, respectively. They describe more than five million entities and contain multiple hundred millions of triples \cite{ringler2017one}. When it comes to relation assertions (RAs), however, we observe -- even for basic properties -- a rather low coverage: More than 50\% of the 1.35 million persons in DBpedia have no birthplace assigned; even more than 80\% of birthplaces are missing in YAGO. At the same time, type assertions (TAs) are not present as well for many instances -- for example, there are about half a million persons in DBpedia not explicitly typed as such \cite{paulheim2013type}.

Missing knowledge in Wikipedia-based knowledge graphs can be attributed to absent information in Wikipedia, but also to the extraction procedures of knowledge graphs. DBpedia uses infobox mappings to extract RAs for individual instances, but it does not explicate any information implicitly encoded in categories. YAGO uses manually defined patterns to assign RAs to entities of matching categories. For example, they extract a person's year of birth by exploiting categories ending with \textit{births}. Consequently, all persons contained in the category \textit{1879 births} are attributed with \textit{1879} as \textit{year of birth} \cite{suchanek2007yago}. Likewise, most existing works, such as \cite{liu2008catriple} and \cite{xu2016learning} leverage textual patterns in the category names.

There are some limitations to such approaches, since, in many cases, very specific patterns are necessary (e.g. \texttt{county,Chester County} for the category \texttt{Townships in Chester County, Pennsylvania}), or the information is only indirectly encoded in the category (e.g. \texttt{timeZone,Eastern\_Time\_Zone} for the same category). In order to capture as much knowledge as possible from categories, we propose an approach that does not learn patterns only from the category names, but exploits the underlying knowledge graph as well.

\begin{figure}[t]
	\centering
	\includegraphics[width=0.9\linewidth]{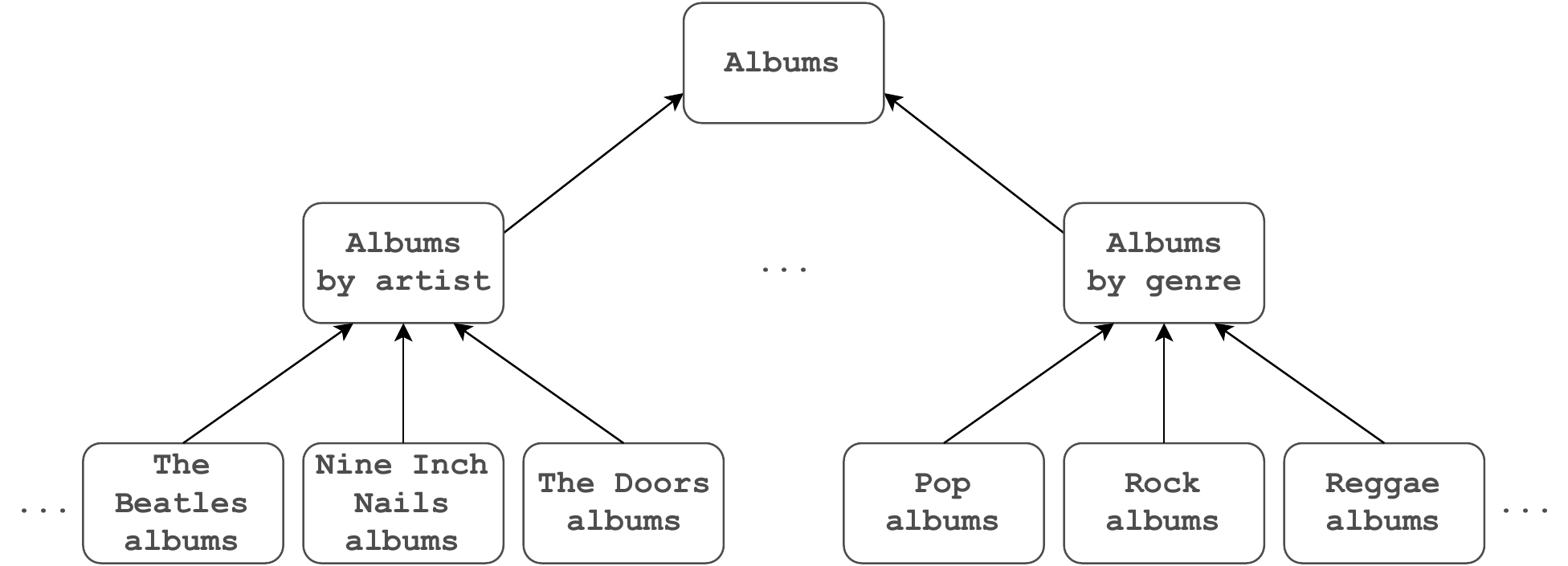}
	\caption{Excerpt of the Wikipedia category graph showing the category \texttt{Albums} together with some of its subcategories.}
	\label{fig:introduction-example}
\end{figure}

While category names are plain strings, we aim at uncovering the semantics in those category names. To that end, we want to extract both type as well as relation information from categories. In the example in Fig.~\ref{fig:introduction-example}, we would, e.g., learn type (1) as well as relation (2-3) axioms, such as:
\begin{eqnarray}
\exists category.\left\{Reggae\_albums\right\} & \sqsubseteq & Album \\
\exists category.\left\{Nine\_Inch\_Nails\_albums\right\} & \sqsubseteq & \exists artist.\left\{Nine\_Inch\_Nails\right\} \\
\exists category.\left\{Reggae\_albums\right\} & \sqsubseteq & \exists genre.\left\{Reggae\right\}
\label{eq:example_axioms}
\end{eqnarray}
Once those axioms are defined, they can be used to fill in missing type and relation assertions for all instances for which those categories have been assigned.

In this paper, we propose the \emph{Cat2Ax} approach to enrich Wikipedia-based knowledge graphs by explicating the semantics in category names. We combine the category graph structure, lexical patterns in category names, and instance information from the knowledge graph to learn patterns in category names (e.g., categories ending in \emph{albums}), and map these patterns to type and relation axioms.
The contributions of this paper are the following:
\begin{itemize}
\item We introduce an approach that extracts axioms for Wikipedia categories using features derived from the instances in a category and their lexicalisations.
\item We extract more than 700k axioms for explicating the semantics of category names at a precision of more than 95\%.
\item Using those axioms, we generate more than 7.7M new assertions in DBpedia at a precision of more than 87\%.
\end{itemize}

The rest of this paper is structured as follows. Section~\ref{relatedwork} frames the approach described in this paper in related works. Section~\ref{preliminaries} lays out the preliminaries of our work, followed by an introduction of our approach in section~\ref{approach}. In section~\ref{experiments}, we discuss an empirical evaluation of our approach. We close with a summary and an outlook on future developments.

\section{Related Work} \label{relatedwork}
With the wider adoption of general purpose knowledge graphs such as DBpedia \cite{lehmann2015dbpedia}, YAGO \cite{mahdisoltani2013yago3}, or Wikidata \cite{vrandevcic2014wikidata}, their quality has come into the focus of recent research \cite{farber2016linked,zaveri2016quality}. The systematic analysis of knowledge graph quality has inspired a lot of research around an automatic or semi-automatic improvement or refinement \cite{paulheim2017knowledge}.

Generally, knowledge graph refinements can be distinguished along various dimensions: the goal (filling missing knowledge or identifying erroneous axioms), the target (e.g., schema or instance level, type or relation assertions, etc.), and the knowledge used (using only the knowledge graph as such or also external sources of knowledge). The approach discussed in this paper extracts axioms on schema level and assertions on instance level using Wikipedia categories as external source of knowledge. 

There are quite a few refinement strategies using additional sources in Wikipedia especially for the extraction of new RAs. Most of them use the text of Wikipedia pages \cite{aprosio2013extending,gerber2011bootstrapping,heist2018language,mintz2009distant}, but also Wikipedia-specific structures, such as tables \cite{munoz2013triplifying,ritze2015matching} or list pages \cite{kuhn2016type,paulheim2013extending}.

For extracting information from categories, there are two signals that can be exploited: (1) lexical information from the category's name, and (2) statistical information of the instances belonging to the category. YAGO, as discussed above, uses the first signal. A similar approach is \emph{Catriple} \cite{liu2008catriple}, which exploits manually defined textual patterns (such as \emph{X by Y}) to identify parent categories which organize instances by objects of a given relation: for example, the category \emph{Albums by genre} has child categories whose instances share the same object for the relation \emph{genre}, and can thus be used to generate axioms such as the one in Equation~3 above. The Catriple approach does not explicitly extract category axioms, but finds 1.27M RAs. A similar approach is taken in \cite{nastase2008decoding}, utilizing POS tagging to extract patterns from category names, but not deriving any knowledge graph axioms from them.

In the area of taxonomy induction, many approaches make use of lexical information when extracting hierarchies of terms. Using Hearst patterns \cite{hearst1992automatic} is one of the most well known method to extract hypernymy relations from text. It has been extended multiple times, e.g., by \cite{kozareva2010learning} who enhance their precision by starting with a set of pre-defined terms and post-filtering the final results. \cite{velardi2013ontolearn} use an optimal branching algorithm to induce a taxonomy from definitions and hypernym relations that have been extracted from text.

The \emph{C-DF} approach \cite{xu2016learning} is an approach of the second category, i.e., it relies on statistical signals. In a first step, it uses probabilistic methods on the category entities to identify an initial set of axioms, and from that, it mines the extraction patterns for category names automatically. The authors find axioms for more than 60k categories and extract around 700k RAs and 200k TAs.

The exploitation of statistical information from category instances is a setting similar to ontology learning \cite{rettinger2012mining}. For example, approaches such as DL-Learner \cite{lehmann2009dl} find description logic patterns from a set of instances. These approaches are very productive when there is enough training data and they provide exact results especially when both positive and negative examples are given. Both conditions are not trivially fulfilled for the problem setting in this paper: many categories are rather small (75\% of categories have fewer than 10 members) and, due to the open world assumption, negative examples for category membership are not given. Therefore, we postulate that both, statistical and lexical information, have to be combined for deriving high-quality axioms from categories.

With Catriple and C-DF, we compare against the two closest approaches in the literature. While Catriple relies solely on lexical information in the category names, and C-DF relies solely on statistical information from the instances assigned to categories, we propose a \emph{hybrid} approach which combines the lexical and statistical signals. Moreover, despite exploiting category names, we do not use any language-specific techniques, so that our approach is in principle language-agnostic.
 
There are other studies using Wikipedia categories for various tasks. Most prominently, taxonomic knowledge graphs such as WiBi \cite{flati2014two} and DBTax \cite{fossati2015unsupervised} are created by cleaning the Wikipedia category graph (which is not an acyclic graph and therefore cannot directly be used as a taxonomy). Implicitly, they also learn type axioms and assertions, but no relation axioms and assertions.

\section{Preliminaries} \label{preliminaries}
The Cat2Ax approach uses three kinds of sources: The Wikipedia category graph, background knowledge from a knowledge graph, and lexicalisations of resources and types in the knowledge graph. In this section, we provide relevant definitions and give background information about the respective sources.

\subsubsection{Wikipedia Categories}
In the version of October 2016,\footnote{We use this version in order to be compatible with the most recent release of DBpedia from October 2016: https://wiki.dbpedia.org/develop/datasets.} the Wikipedia category graph contains 1,475,015 categories that are arranged in a directed, but not acyclic graph, although often referred to as a \emph{category hierarchy}. This graph does not only contain categories used for the categorisation of content pages, but also ones that are used for administrative purposes. We follow an approach similar to \cite{ponzetto2007deriving} and use only categories below \texttt{Main topic classifications} while also getting rid of categories having one of the following words in their name: \textit{wikipedia, lists, template, stub}. This leaves us with 1,299,665 categories.

\subsubsection{Background Knowledge}
As background knowledge, our approach requires a knowledge graph \textit{KG} that is based on Wikipedia. The knowledge graph is comprised of a set of \textit{resources} which are connected by \textit{relations}, and an \textit{ontology} which defines their classes, interrelations, and restrictions of usage. A resource in the knowledge graph describes exactly one article in Wikipedia. When we are referring to DBpedia in our examples and experiments, we use the prefix \textit{dbr:} for resources and \textit{dbo:} for properties and types.

With \textit{resources(c)} we refer to the set of resources with a corresponding article assigned to the category \textit{c}. To get an estimate of how likely a combination of a property $p$ and a value $v$ occurs within the resources of a category $c$, we calculate their frequencies using background knowledge from the knowledge graph $KG$:
\begin{equation}
freq(c,p,v) = \frac{|\left\{r | r \in resources(c) \land (r,p,v) \in KG\right\}|}{|resources(c)|}
\end{equation}
For $p = $ \texttt{rdf:type}, we compute type frequencies of $c$.

\textbf{Example 1} The category \texttt{The Beatles albums} has 24 resources, 22 of which have the type \texttt{dbo:Album}. This results in a type frequency $freq($\texttt{The Beatles albums, rdf:type, dbo:Album}$)$ of 0.92.
\\\\
For $p$ being any other property of $KG$, we compute relation frequencies of $c$.

\textbf{Example 2} Out of the 24 resources of \texttt{The Beatles albums}, 11 resources have \texttt{dbr:Rock\_and\_roll} as \texttt{dbo:genre}, resulting in a relation frequency $freq($\texttt{The Beatles albums, dbo:genre, dbr:Rock\_and\_roll}$)$ of 0.46.

\subsubsection{Resource/Type Lexicalisations}
A lexicalisation is a word or phrase used in natural language text that refers to a resource or type in the knowledge graph. For an entity \textit{e}, \textit{lex(e)} contains all its lexicalisations, and \textit{lexCount(e,l)} is the count of how often a lexicalisation \textit{l} has been found for \textit{e}. When the count of a lexicalisation \textit{l} is divided by the sum of all counts of lexicalisations for an entity \textit{e}, we have an estimate of how likely \textit{e} will be expressed by \textit{l}.

We are, however, interested in the inverse problem: Given a lexicalisation \textit{l}, we want the probability of it expressing an entity \textit{e}. We define $lex^{-1}(l)$ as the set of entities having \textit{l} as lexicalisation. The lexicalisation score -- the probability of an entity \textit{e} being expressed by the lexicalisation \textit{l} -- is then computed by the fraction of how often \textit{l} expresses \textit{e} compared to all other entities:
\begin{equation}
lexScore(e,l) = \frac{lexCount(e,l)}{\sum_{e' \in lex^{-1}(l)}{lexCount(e',l)}}
\end{equation}

\textbf{Example 3} We encounter the word \texttt{lennon} in Wikipedia and want to find out how likely it is that the word refers to the resource \texttt{dbr:John\_Lennon}, i.e. we compute $lexScore($\texttt{dbr:John\_Lennon, lennon}$)$. In total, we have 357 occurrences of the word for which we know the resource it refers to. 137 of them actually refer to \texttt{dbr:John\_Lennon}, while others refer, e.g., to the soccer player \texttt{dbr:Aaron\_Lennon} (54 times) or \texttt{dbr:Lennon,\_Michigan} (14 times). We use the occurrence counts to compute a $lexScore($\texttt{dbr:John\_Lennon, lennon}$)$ of 0.42.

We compute lexicalisation scores for both resources and types in our experiments with DBpedia. The lexicalisations of resources are already provided by DBpedia \cite{bryl2015gathering}. They are gathered by using the anchor texts of links between Wikipedia articles. For types, however, there is no such data set provided.

To gather type lexicalisations from Wikipedia, we apply the following methodology: For every type \textit{t} in the DBpedia ontology, we crawl the articles of all resources having type \textit{t} and extract hypernymy relationships using Hearst patterns \cite{hearst1992automatic}. To ensure that we are only extracting relationships for the correct type, we use exclusively the ones having a lexicalisation of the page's resource as their subject. To increase the coverage of type lexicalisations, we intentionally do not count complete phrases, but individual words of the extracted lexicalisation. For the calculation of the lexicalisation scores of a phrase, we simply sum up the counts of the phrase's words.

\textbf{Example 4} We extract lexicalisations for the type \texttt{dbo:Band}. The resource \texttt{dbr:Nine\_Inch\_Nails} has the appropriate type, hence we extract hypernymy relationships in its article text. In the sentence \textit{"Nine Inch Nails is an American industrial rock band [..]"} we find the subject \textit{Nine Inch Nails} and the object \textit{American industrial rock band}. As the subject is in lex(\texttt{dbr:Nine\_Inch\_Nails}), we accept the object as lexicalisation of \texttt{dbo:Band}. Consequently, the lexicalisation count of the words \textit{American, industrial, rock, band} is increased by one, and, for each of those words encountered, the lexicalisation score for the class \texttt{dbo:Band} increases.

\section{Approach} \label{approach}
\begin{figure}[t]
	\centering
	\includegraphics[width=\linewidth]{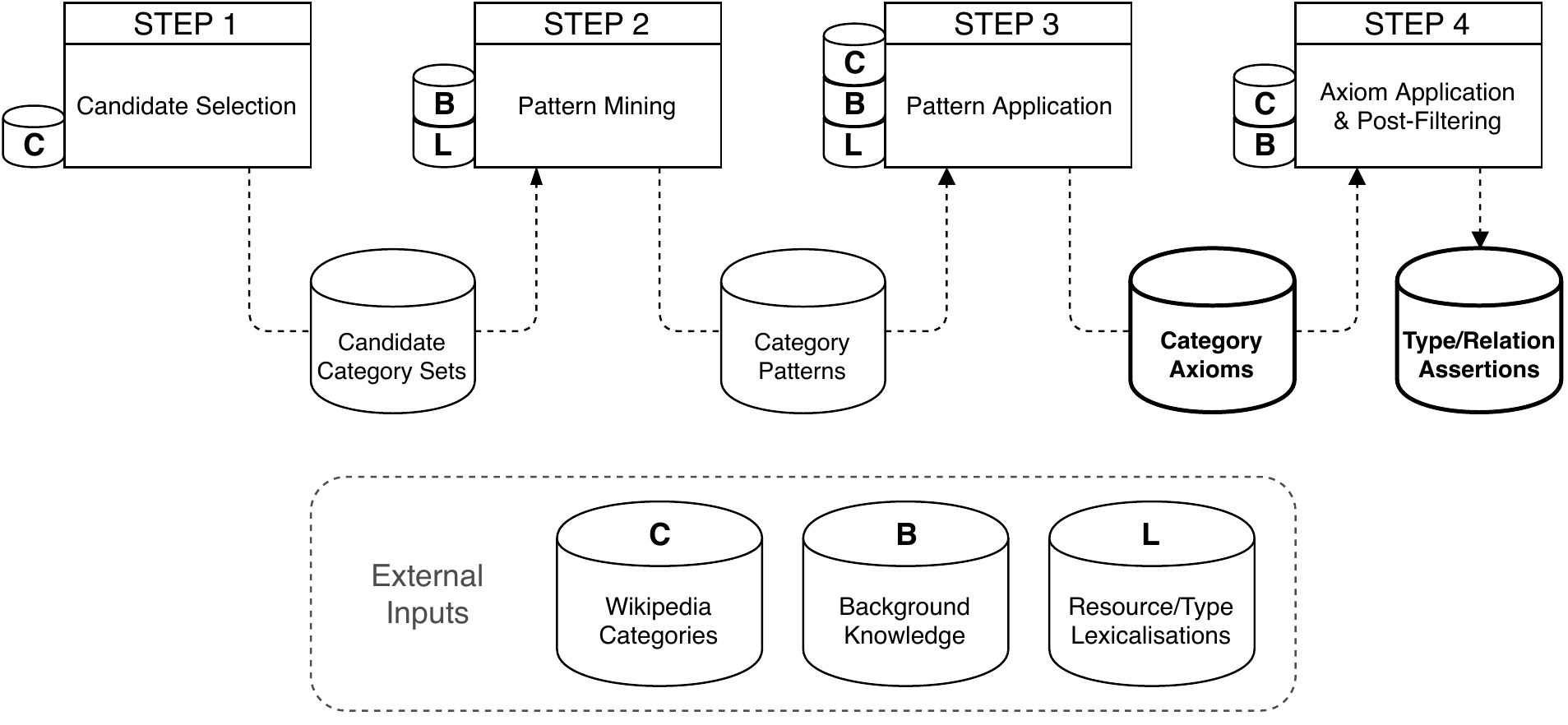}
	\caption{Overview of the Cat2Ax approach.}
	\label{fig:approach-overview}
\end{figure}
The overall approach of Cat2Ax is shown in Fig.~\ref{fig:approach-overview}. The external inputs have already been introduced in Section~\ref{preliminaries}. The outputs of the approach (marked in bold font) are twofold: A set of axioms which define restrictions for resources in a category and thus can be used to enhance an ontology of a knowledge graph, and a set of assertions which are novel facts about resources in the graph.

The approach has four major steps: The \textit{Candidate Selection} uses hierarchical relationships in the Wikipedia category graph to form sets of categories that are likely to share a property that can be described by a textual pattern.

In the \textit{Pattern Mining} step, we identify such patterns in the names of categories that are characteristic for a property or type. To achieve that, we use background knowledge about resources in the respective categories as well as lexicalisations. Furthermore, we promote a pattern only if it applies to a majority of the categories in a candidate set. 

In the \textit{Pattern Application} step, we apply the extracted patterns to all categories in order to find category axioms. Here, we again rely on background knowledge and lexicalisations for the decision of whether a pattern is applicable to the category.

Finally, we generate assertions by applying the axioms of a category to its resources and subsequently use post-filtering to remove assertions that would create contradictions in the knowledge graph.

\subsection{Candidate Selection} \label{candidateselection}
In this first step, we want to extract sets of categories with names that indicate a shared relation or type. We base the extraction of such candidate category sets on two observations: 

The first one is inspired by the Catriple approach \cite{liu2008catriple}. They observed that in a parent-child relationship of categories, the parent often organizes its children according to a certain property. Contrary to Catriple, we do not use the parent category to identify this property, but we rather use the complete set of children to find their similarities and differences.

As we now know from the first observation, the children of a category can have certain similarities (which are the reason that they have the same parent category) and differences (which are the reason that the parent was split up into child categories). As a second observation, we discovered that, when the children of a category are organized by a certain property, their names have a shared part (i.e. a common prefix and/or postfix) and a part that differs for each category. We found that the shared part is often an indicator for the type of resources that are contained in the category, while the differing part describes the value of the property by which the categories are organized.

Using these observations, we produce the candidate category sets by looking at the children of each Wikipedia category and forming groups out of children that share a prefix and/or postfix.

\textbf{Example 5} In Fig.~\ref{fig:introduction-example}, we see parts of two candidate category sets that both have the postfix \textit{albums}. The first one contains 143 children of the category \texttt{Albums by artist}. The second one contains 45 children of the category \texttt{Albums by genre}.

Note that we sometimes form multiple candidate category sets from category's children as there might be more than one shared pre- or postfix.

\textbf{Example 6}  The children of the category \texttt{Reality TV participants} yield three candidate sets ending on \textit{participants}, \textit{contestants}, and \textit{members}.

\subsection{Pattern Mining} \label{patternmining}
For each of the candidate category sets, we want to discover a characteristic property and type. Therefore, we identify patterns that will be used in the subsequent steps to extract category axioms. Each of the patterns consists of a textual pattern (i.e. the shared part in the names of categories) and the implication (i.e. the shared property or type).

To determine the characteristic property, we inspect every individual category in the candidate set and compute a score for every possible relation in the category. As mentioned in Section~\ref{candidateselection}, the value of a relation differs for the categories in a set. We thus focus on finding the property with the highest score and disregard relation values. To that end, we aggregate the scores from all categories and choose the property that performs best over the complete category set. For this property, we learn a pattern that covers the complete candidate category set.

The score of a relation \textit{(p, v)} for a category \textit{c} consists of two parts with one being based on background knowledge and the other on lexical information. The latter uses the part $c_{var}$ of a category's name that differs between categories in the set to compute an estimate of how likely $c_{var}$ expresses the value of the relation. The score is computed as follows:
\begin{equation}
score_{rel}(c,p,v) = freq(c,p,v) * lexScore(v, c_{var})
\end{equation}
Note that \textit{freq(c,p,v)} is only greater than zero for relations of the resources in \textit{resources(c)} which drastically reduces the amount of scores that have to be computed.

\textbf{Example 7} For the category \texttt{The Beatles albums}, we compute an individual relation score for each property-value pair in $KG$ having a resource in $resources($\texttt{The Beatles albums}$)$ as their subject. To compute, e.g., $score_{rel}($\texttt{The Beatles albums, dbo:artist, dbr:The\_Beatles}$)$, we multiply the frequency $freq($\texttt{The Beatles albums, dbo:artist, dbr:The\_Beatles}$)$ with the lexicalisation score $lexScore($\texttt{dbr:The\_Beatles, The Beatles}$)$.

As an aggregation function for the scores we use the median. Heuristically, we found that the property with the highest median of scores is suited to be the characteristic property for a category set. To avoid learning incorrect patterns, we discard the property if it cannot be found in at least half of the categories in the set, i.e,. if the median of scores is zero.

\textbf{Example 8} After computing all the relation scores for all categories in the category set formed by the 143 children of \texttt{Albums by artist}, we aggregate the computed scores by their property and find \texttt{dbo:artist} to have the highest median score.

The support of a pattern is the count of how often a pattern has been learned for a category. If we discover a valid property for a category set, the support of the respective property pattern is increased by the number of categories in the set. We assume hereby that, if this property is characteristic for the majority of categories, then it is characteristic for all categories in the set.

For the extraction of characteristic types we apply the exact same methodology, except for the calculation of the score of a type. We compute the score of a type \textit{t} in the category \textit{c} using its frequency in $c$ and a lexicalisation score derived from the shared part $c_{fix}$ in a category's name:
\begin{equation}\label{eq:relscore}
score_{type}(c,t) = freq(c,\texttt{rdf:type},t) * lexScore(t, c_{fix})
\end{equation}

\textbf{Example 9} For the category sets formed by the children of \texttt{Albums by artist} and \texttt{Album by genre} in Fig.~\ref{fig:introduction-example}, we find the following property patterns to have the highest median scores:
\begin{itemize}
\item $PP_1$ = "$<$lex(\textit{dbr:res})$>$ albums" $\sqsubseteq \exists$\texttt{dbo:artist}.\{\textit{dbr:res}\} \\
\item $PP_2$ = "$<$lex(\textit{dbr:res})$>$ albums" $\sqsubseteq \exists$\texttt{dbo:genre}.\{\textit{dbr:res}\}
\end{itemize}
We increase the support of $PP_1$ by 143 and $PP_2$ by 45. For both sets, we extract the type pattern $TP_1$ = "$<$lex(\textit{dbr:res})$>$ albums" $\sqsubseteq$ \texttt{dbo:Album} and increase its support by 188 (respectively using the counts from Example~5).

\subsection{Pattern Application}
Before we can apply the patterns to the categories in Wikipedia to identify axioms, we need to define a measure for the confidence of a pattern. This is especially necessary because, as shown in Example~9, we can find multiple implications for the same textual pattern. We define the confidence $conf(P)$ of a pattern \textit{P} as the quotient of the support of \textit{P} and the sum of supports of all the patterns matching the same textual pattern as \textit{P}.

\textbf{Example 10} Assuming $PP_1$ and $PP_2$ of Example~9 are the only property patterns that we found, we have a pattern confidence of 0.76 for $PP_1$ and 0.24 for $PP_2$.

Next, we apply all our patterns to the categories of Wikipedia and compute an axiom confidence by calculating the fit between the category and the pattern. Therefore, we can reuse the scores from Equations~6-7 and combine them with the confidence of the pattern. As a relation pattern only specifies the property of the axiom, we compute the axiom confidence for every possible value of the axiom's property in order to have a ranking criterion. For a category \textit{c}, a property pattern \textit{PP} with property $p_{PP}$ and a possible value \textit{v}, we compute the confidence as follows:
\begin{equation}
conf(c,PP,v) = conf(PP) * score_{rel}(c,p_{PP},v)
\end{equation}
And similarly, for a type pattern \textit{TP} with type $t_{TP}$:
\begin{equation}
conf(c,TP) = conf(TP) * score_{type}(c,t_{TP})
\end{equation}

Using the confidence scores, we can control the quality of extracted axioms by only accepting those with a confidence greater than a threshold $\tau$. To find a reasonable threshold, we will inspect and evaluate the generated axioms during our experiments.

\textbf{Example 11} Both patterns, $PP_1$ and $PP_2$, from Example~9 match the category \texttt{Reggae albums}. Using $PP_1$, we can not find an axiom for the category as there is no evidence in DBpedia for the property \texttt{dbo:artist} together with any resources that have the lexicalisation \textit{Reggae} (i.e. $score_{rel}$ is equal to 0). For $PP_2$, however, we find the axiom \texttt{(Reggae albums, dbo:genre, dbr:Reggae)} with a confidence of 0.18.

For a single category, multiple property or type patterns can have a confidence greater than $\tau$. The safest variant for property and type patterns is to accept only the pattern with the highest confidence and discard all the others. But we found that multiple patterns can imply valid axioms for a category and thus follow a more differentiated selection strategy.

For relation axioms, we accept multiple axioms as long as they have different properties. When the properties are equal, we accept only the axiom with higher confidence.

\textbf{Example 12} For the category \texttt{Manufacturing companies established in 1912} (short: $c_1$), we find the axioms ($c_1$, \texttt{dbo:foundingYear}, 1912) and ($c_1$, \texttt{dbo:industry}, \texttt{dbr:Manufacturing}). As they have different properties, we accept both.

\textbf{Example 13} For the category \texttt{People from Nynäshamn Municipality} (short: $c_2$), we find the axioms ($c_2$, \texttt{dbo:birthPlace}, \texttt{dbr:Nynäshamn\_Municipality}) and ($c_2$, \texttt{dbo:birthPlace}, \texttt{dbr:Nynäshamn}). As they have the same property, we only accept the former as its confidence is higher.

For type axioms, we accept the axioms with the highest confidence and any axioms with a lower confidence that imply sub-types of the already accepted types.

\textbf{Example 14} For the category \texttt{Missouri State Bears baseball coaches} (short: $c_3$), we find the axioms ($c_3$, \texttt{rdf:type}, \texttt{dbo:Person}) and ($c_3$, \texttt{rdf:type}, \texttt{dbo:CollegeCoach}). Despite the latter having a lower confidence than the former, we accept both because \texttt{dbo:CollegeCoach} is a sub-type of \texttt{dbo:Person}.

\subsection{Axiom Application and Post-Filtering}
With the category axioms from the previous step, we generate new assertions by applying the axiom to every resource of the category.

\textbf{Example 15} We apply the axiom (\texttt{Reggae albums,dbo:genre,dbr:Reggae}) to all resources of \texttt{Reggae albums} and generate 50 relation assertions, 13 of which are not yet present in DBpedia.

Categories can contain special resources that do not actually belong to the category itself but, for example, describe the topic of the category. The category \texttt{Landforms of India}, for example, contains several actual landforms but also the resource \texttt{Landforms of India}. To avoid generating wrong assertions for such special resources, we filter all generated assertions using the existing knowledge in the knowledge base.

For relation assertions, we use the functionality of its property to filter invalid assertions. Accordingly, we remove a relation assertion \textit{(s, p, o)} if the property \textit{p} is functional\footnote{Since the DBpedia ontology does not define any functional object properties, we use a heuristic approach and treat all properties which are used with multiple objects on the same subject in less than 5\% of the subjects as functional. This heuristic marks 710 out of 1,355 object properties as functional.} and there is an assertion \textit{(s, p, o')} with $o \neq o'$ already in the knowledge base.

\textbf{Example 16} Out of the 13 new \texttt{dbo:genre} axioms generated for the category \texttt{Reggae albums} in the previous example, nine refer to resources which do not have a \texttt{dbo:genre} at all, and four add a genre to a resource which already has one or more values for \texttt{dbo:genre}. The latter is possible since \texttt{dbo:genre} is not functional.

\textbf{Example 17} The relation assertion (\texttt{dbr:Bryan\_Fisher, dbo:birthYear}, 1982) is removed because DBpedia contains the triple (\texttt{dbr:Bryan\_Fisher, dbo:birthYear}, 1980) already, and \texttt{dbo:birthYear} is functional.

To identify invalid type assertions, we use the disjointness axioms of the ontology of the knowledge base, and remove any type assertion that, if added to the knowledge base, would lead to a conflict of disjointness.

\textbf{Example 18} The assertion (\texttt{dbr:Air\_de\_Paris, rdf:type, dbo:Person}) is removed because the subject has already the type \texttt{dbo:Place}, which is disjoint with \texttt{dbo:Person}.

\section{Experiments} \label{experiments}
In this section, we first provide details about the application of the Cat2Ax approach with DBpedia as background knowledge. Subsequently, we discuss the evaluation of Cat2Ax and compare it to the related approaches. For the implementation of the approaches we used the Python libraries spaCy\footnote{\url{https://spacy.io/}} and nltk\footnote{\url{https://www.nltk.org/}}. The code of Cat2Ax\footnote{https://github.com/nheist/Cat2Ax} and all data\footnote{http://data.dws.informatik.uni-mannheim.de/Cat2Ax} of the experiments are freely available.

\subsection{Axiom Extraction using DBpedia} \label{axiomextractionusingdbpedia}
The following results are extracted using the most recent release of DBpedia.\footnote{Release of October 2016: https://wiki.dbpedia.org/develop/datasets}

\textbf{Candidate Selection} We find 176,785 candidate category sets with an average size of eight categories per set. From those sets, 60,092 have a shared prefix, 76,791 a shared postfix, and 39,902 both a shared prefix and postfix.

\textbf{Pattern Mining} We generate patterns matching 54,465 different textual patterns. For 24,079 of them we imply properties, for 54,096 we imply types. On average, a property pattern implies 1.22 different properties while a type pattern implies 1.08 different types. Table~\ref{tab:example-patterns} lists exemplary patterns that match a prefix (rows 1-2), a postfix (rows 3-4), and both a prefix and a postfix (rows 5-6).

{\renewcommand{\arraystretch}{1.25}%
	\begin{table}[t]
		\centering
		\begin{tabular}{| C{0.5cm} | C{4.9cm} | C{4.4cm} | C{0.85cm} | C{0.85cm} |}
			\hline
			& Textual pattern & Implication & Sup. & Conf.\\
			\hline \hline
			1 & Films directed by $<$lex(\textit{dbr:res})$>$ & $\sqsubseteq \exists$dbo:director.\{\textit{dbr:res}\} & 7661 & 1.00 \\
			\hline
			2 & Films directed by $<$lex(\textit{dbr:res})$>$ & $\sqsubseteq$ dbo:Film & 7683 & 1.00 \\
			\hline
			\makecell{\\3\\\\} & \makecell{\\$<$lex(\textit{dbr:res})$>$ albums\\\\} & \makecell{$\sqsubseteq \exists$dbo:artist.\{\textit{dbr:res}\}\\ $\sqsubseteq \exists$dbo:genre.\{\textit{dbr:res}\}\\ $\sqsubseteq \exists$dbo:recordLabel.\{\textit{dbr:res}\}} & \makecell{31426\\552\\411} & \makecell{0.97\\0.02\\0.01} \\
			\hline
			4 & $<$lex(\textit{dbr:res})$>$ albums & $\sqsubseteq$ dbo:Album & 33542 & 1.00 \\
			\hline
			5 & Populated places in $<$lex(\textit{dbr:res})$>$ district & \makecell{$\sqsubseteq \exists$dbo:isPartOf.\{\textit{dbr:res}\}\\ $\sqsubseteq \exists$dbo:district.\{\textit{dbr:res}\}} & \makecell{269\\51} & \makecell{0.84\\0.16} \\
			\hline
			6 & Populated places in $<$lex(\textit{dbr:res})$>$ district & $\sqsubseteq$ dbo:Settlement & 362 & 1.0 \\
			\hline
		\end{tabular}
		\caption{Examples of discovered textual patterns and possible implications.}
		\label{tab:example-patterns}
	\end{table}
}

\begin{figure}[t]
	\centering
	\includegraphics[width=0.75\linewidth]{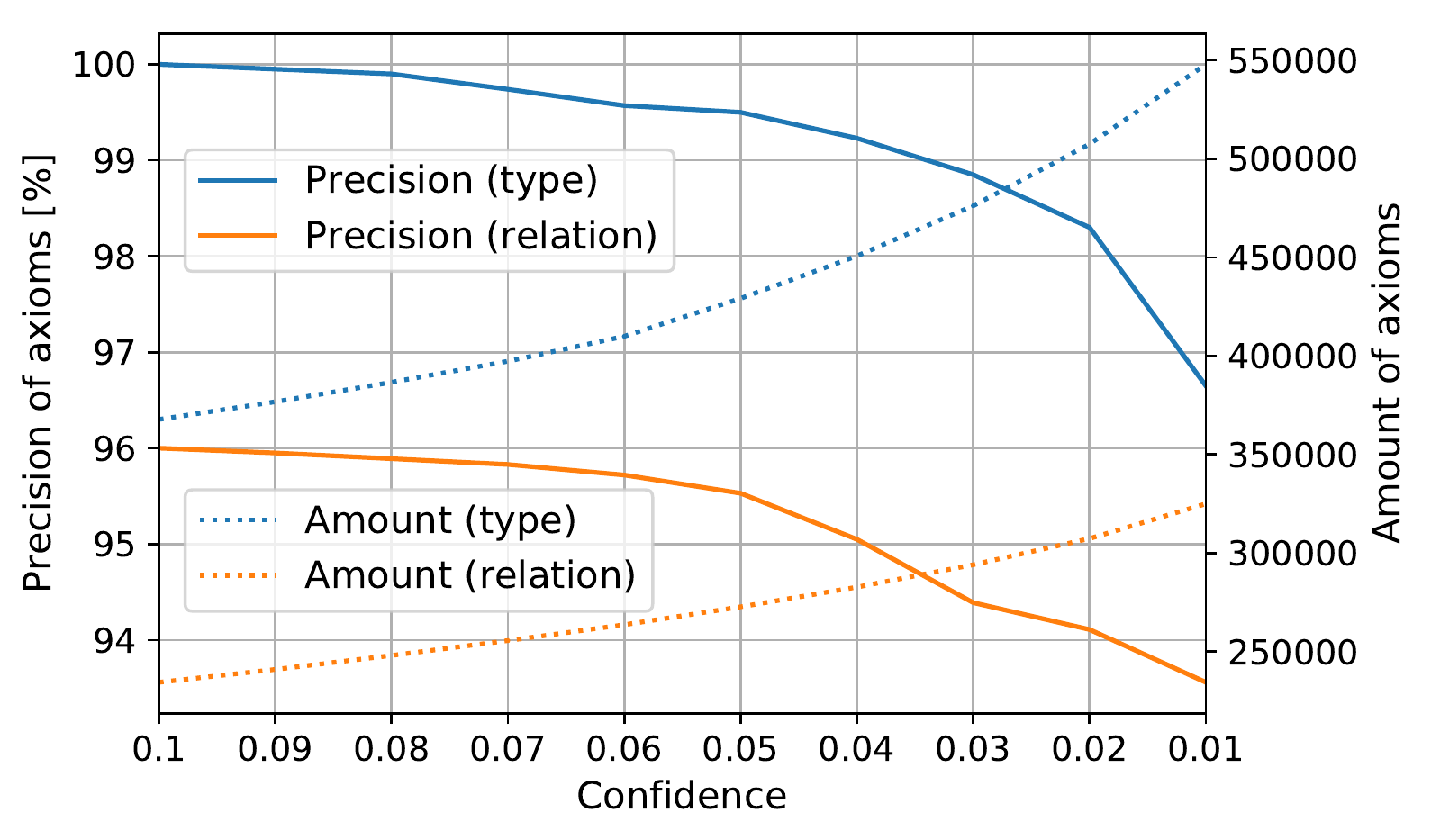}
	\caption{Performance of the pattern application for varying confidence intervals. The precision values have been determined by the authors by manual evaluation of 50 examples per interval.}
	\label{fig:min-confidence}
\end{figure}

\textbf{Pattern Application} We have to determine a threshold $\tau$ for the minimum confidence of an accepted axiom. Therefore, we have sampled 50 generated axioms for 10 confidence intervals each ($[0.01,0.02)$,$[0.02,0,03)$, ..., $[0.09,0.10)$ and $[0.10,1.00]$), and manually evaluated their precision. The results are shown in Figure~\ref{fig:min-confidence}. We can observe that the precision considerably drops for a threshold lower than $\tau = 0.05$, i.e., for those axioms which have a confidence score less than 5\%. Hence, we choose $\tau = 0.05$ for a reasonable balance of axiom precision and category coverage.

With a confidence threshold $\tau$ of 0.05, we extract 272,707 relation axioms and 430,405 type axioms. In total, they cover 501,951 distinct Wikipedia categories. 

\textbf{Axiom Application and Post-Filtering} Applying the extracted axioms to all Wikipedia categories results in 4,424,785 relation assertions and 1,444,210 type assertions which are not yet contained in DBpedia. For the type assertions, we also compute the transitive closure using the \texttt{rdfs:subclassOf} statements in the ontology (e.g., also asserting \texttt{dbo:MusicalWork} and \texttt{dbo:Work} for a type axiom learned for type \texttt{dbo:Album}), and thereby end up with 3,342,057 new type assertions (excluding the trivial type \texttt{owl:Thing}).

Finally, we remove 72,485 relation assertions and 15,564 type assertions with our post-filtering strategy. An inspection of a small sample of the removed assertions shows that approximately half of the removed assertions are actually incorrect.

\subsection{Comparison with Related Approaches}

We compare Cat2Ax with the two approaches that also use Wikipedia categories to learn axioms and/or assertions for DBpedia: Catriple \cite{liu2008catriple} and C-DF \cite{xu2016learning}. As both of them use earlier versions of DBpedia and there is no code available, we re-implemented both approaches and run them with the current version in order to have a fair comparison. For the implementation, we followed the algorithm descriptions in their papers and used the variant with the highest reported precision (i.e., for Catriple, we do not materialize the category hierarchy, and for C-DF, we do not apply patterns iteratively). Running Cat2Ax, Catriple, and C-DF with DBpedia takes 7, 8, and 12 hours, respectively.

Table~\ref{tab:approach-comparison} shows the extraction and evaluation results of the three approaches. For both kinds of axioms and assertions, we evaluate 250 examples per approach. Since the Catriple approach does not produce type information, this adds up to a total of 2,500 examples (1,250 axioms and 1,250 assertions). Each example is labeled by three annotators from the crowdsourcing marketplace Amazon Mechanical Turk.\footnote{\url{https://www.mturk.com/}} For the labeling, the axioms and assertions are presented in natural language (using labels from DBpedia) and have to be annotated as being either correct or incorrect. The annotators evaluate batches of 50 examples which are selected from the complete example pool and displayed in a random order. The inter-annotator agreement according to Fleiss' kappa \cite{fleiss_kappa} is 0.54 for axioms and 0.53 for assertions which indicates moderate agreement according to \cite{fleiss_interpretation}.

{\renewcommand{\arraystretch}{1.25}%
	\begin{table}[t]
		\centering
		\begin{tabular}{| C{2.3cm} | C{2.3cm} | C{2.3cm} | C{2.3cm} | C{2.3cm} |}
			\hline
			\textbf{Approach} & \textbf{Count} & \textbf{Precision [\%]} & \textbf{Count} & \textbf{Precision [\%]}\\
			\hline \hline
			& \multicolumn{2}{c|}{Relation axioms} & \multicolumn{2}{c|}{Type axioms}\\
			\hline \hline
			Cat2Ax & 272,707 & 95.6 & 430,405 & 96.8 \\
			\hline
			C-DF & 143,850 & 83.6 & \hphantom{0}28,247 & 92.0 \\
			\hline
			Catriple & 306,177 & 87.2 & -- & -- \\
			\hline\hline
			& \multicolumn{2}{c|}{Relation assertions} & \multicolumn{2}{c|}{Type assertions}\\
			\hline \hline
			Cat2Ax & 4,424,785 (7,554,980) & 87.2\newline(92.1) & \hphantom{0}3,342,057 (12,111,194) & 90.8\newline(95.7) \\
			\hline
			C-DF & \hphantom{0,}766,921 (2,856,592) & 78.4\newline(93.4)& \hphantom{0,}198,485 (2,352,474) & 76.8\newline(97.1) \\
			\hline
			Catriple & 6,260,972 (6,836,924) & 74.4\newline(76.5) & -- & -- \\
			\hline
		\end{tabular}
		\caption{Total number of axioms/assertions and precision scores, based on the crowd-sourced evaluation. Numbers in parentheses denote the \emph{total} number of assertions generated (including those already existing in DBpedia), as well as the precision estimation of those total numbers. The latter were derived as a weighted average from the human annotations and the overall correctness of existing assertions in DBpedia according to \cite{farber2016linked}.}
		\label{tab:approach-comparison}
	\end{table}
}

In comparison with existing approaches, Cat2Ax outperforms C-DF both in quality and quantity of the created axioms. Catriple produces about 40\% more relation assertions, but at a considerably lower precision, and is not able to generate type axioms and assertions.

Despite our efforts of post-filtering generated assertions, a large gap between the precision of axioms and assertions can be observed. This is more evident when looking at \emph{new} assertions, while the overall precision considering both kinds of assertions, which are in DBpedia and which are not, is typically higher. Moreover, there is a small number of axioms which are incorrect and at the same time very productive, i.e., they contribute a lot of new incorrect assertions. To further look into these issues, we manually inspected some of those axioms and identified three major causes of errors:

\textbf{Incorrect data in DBpedia} We extract the axiom (\texttt{Roads on the National Register of Historic Places in Arizona, rdf:type, dbo:Building}) because many roads in DBpedia are typed as buildings.

\textbf{Correlation instead of causation} We extract the axiom (\texttt{University of Tabriz alumni, dbo:birthPlace, dbr:Tabriz}) because people often study in the vicinity of their birthplace.

\textbf{Incorrect generalisation} We extract the axiom (\texttt{Education in Nashik district, rdf:type, dbo:University}), which holds for many instances in the category, but not for all of them. This kind of error is most often observed for mixed categories -- as in the example, the category contains both universities and schools.

\begin{figure}[t]
	\begin{subfigure}{.47\textwidth}
		\centering
		\includegraphics[width=\linewidth]{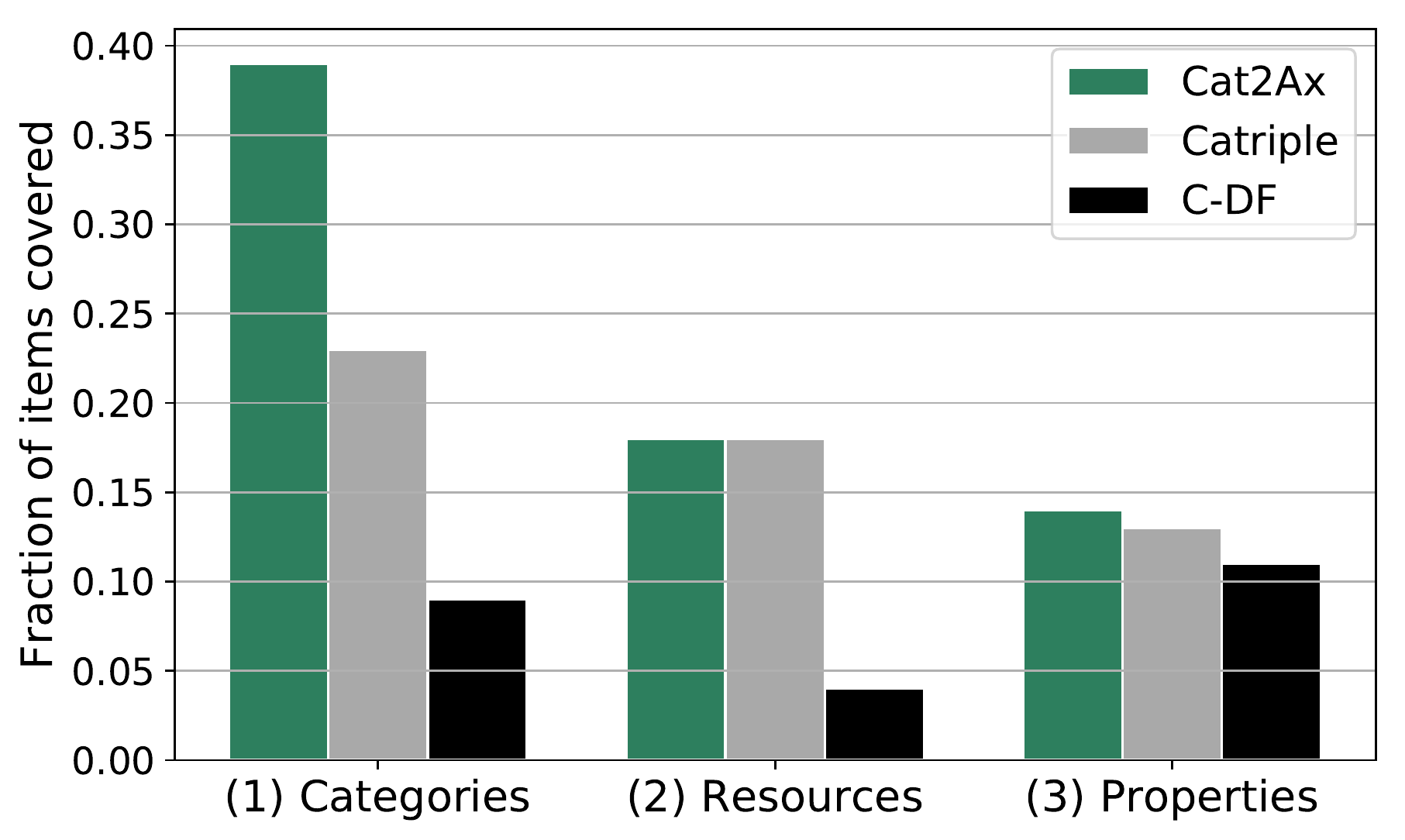}
		\caption{Fraction of (1) categories with at least one axiom, (2) resources with at least one assertion, (3) properties with at least 100 assertions.}
		\label{fig:dbpedia-coverage}
	\end{subfigure}\hfill
	\begin{subfigure}{.5\textwidth}
		\centering
		\includegraphics[width=\linewidth]{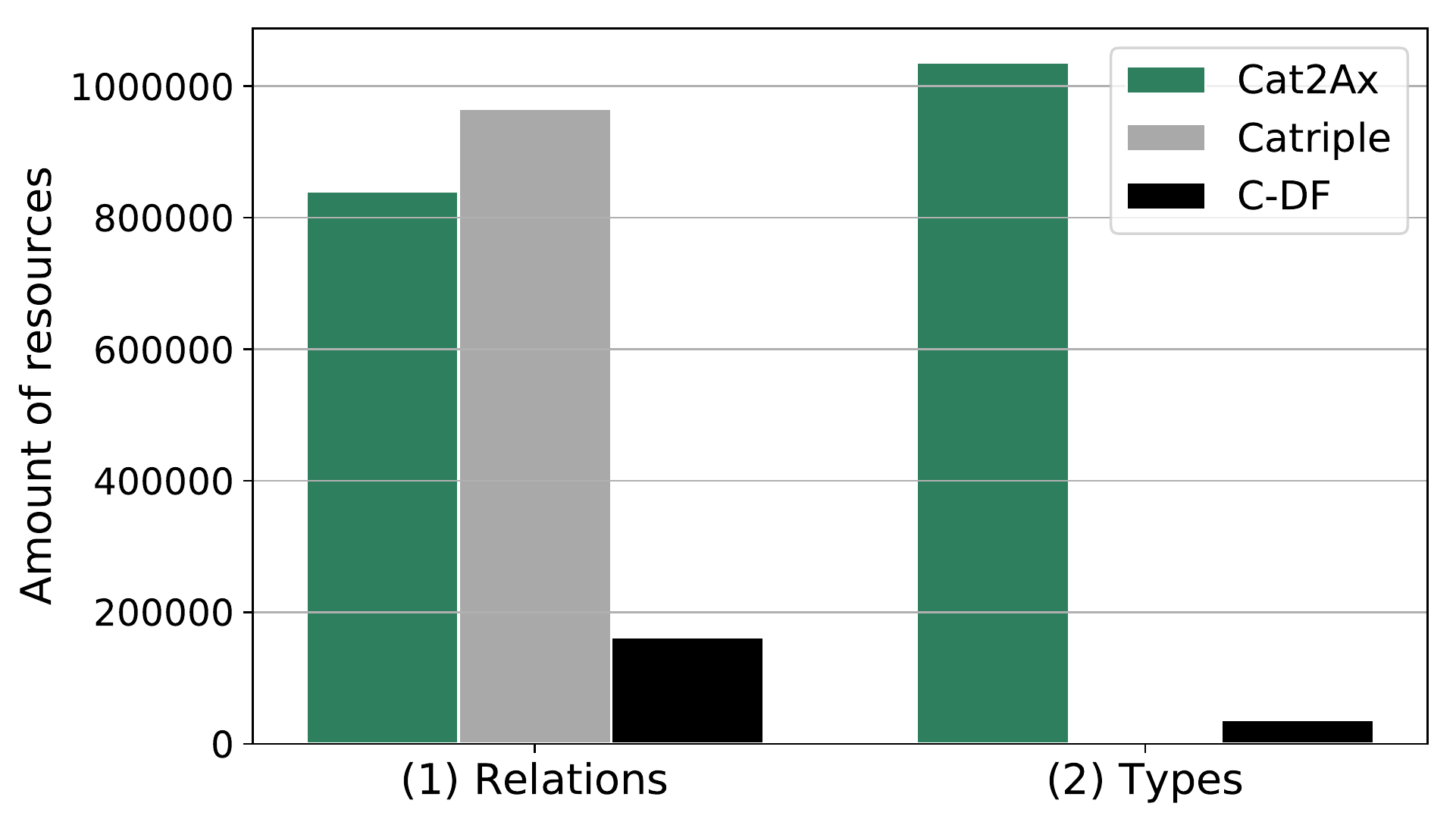}
		\caption{Number of resources without assertions in DBpedia for which (1) a relation assertion or (2) type assertion has been found. \hphantom{placeholder for equal height captions}}
		\label{fig:dbpedia-unknown-resources}
	\end{subfigure}
	\caption{Comparison of the extracted results.}
	\label{fig:approach-comparison}
\end{figure}

In Fig.~\ref{fig:approach-comparison} we compare the results of the three approaches regarding their coverage of DBpedia. Fig.~\ref{fig:dbpedia-coverage} shows the number of covered (1) categories, (2) resources, and (3) properties. At (1) we see that Cat2Ax finds an axiom for almost 40\% of Wikipedia's categories. The difference between Cat2Ax and Catriple is, however, not visible in (2) anymore. This can be traced back to Catriple not using any background knowledge during their creation of results and thus producing axioms that are more productive in terms of generated assertions. Furthermore, (3) shows that all approaches find assertions for a comparable number of properties.

Fig.~\ref{fig:dbpedia-unknown-resources} shows statistics for resources that are currently not described by any relation or type in DBpedia. While Cat2Ax and Catriple both find relations for almost one million resources, Cat2Ax additionally finds types for more than one million untyped resources.

\section{Conclusion} \label{conclusion}
In this paper, we have presented an approach that extracts high-quality axioms for Wikipedia categories. Furthermore, we used the axioms to mine new assertions for knowledge graphs. For DBpedia, we were able to add 4.4M relation assertions at a precision of 87.2\% and 3.3M type assertions at a precision of 90.8\%. Our evaluation showed that we produce significantly better results than state-of-the-art approaches.

So far, we have only considered direct assignments to categories. Exploiting the containment relations between categories and materialising the category assignments would help the approach in two respects -- the extraction of axioms is supported by more precise relation and type frequencies, and the extracted axioms can be applied to a larger number of resources, leading to a higher number of generated assertions. However, this materialisation is not straightforward as the Wikipedia category graph is not acyclic. Currently, we are working on extracting a proper hierarchy from the Wikipedia category graph, which can then be used as a basis for a refined approach.

Moreover, we currently consider all the generated patterns in isolation. But we plan to combine patterns on two dimensions. Firstly, we want to investigate methods to form more generalised patterns out of the currently extracted ones. We expect this to improve the quality of pattern confidence values and the patterns are applicable to more categories. Secondly, property and type patterns and their generated axioms can be combined to provide a better post-filtering of assertions. Given that we know that a relation axiom and a type axiom belong together, and we encounter a single inconsistency in their set of generated axioms, we can discard the complete set.

In previous works, the exploitation of list pages has been discussed for learning new type and relation assertions for instances \cite{kuhn2016type,paulheim2013extending}. We plan to extend the approach in this paper to list pages as well. To that end, we need to robustly extract entities from a list page (which is not straightforward since not all links on a list page necessarily link to entities of the corresponding set), and we need to allocate a list page to a position in the category graph.

It is important to note that, although we carried out experiments with DBpedia, the approach is not limited to only this knowledge graph. Any knowledge graph linked to Wikipedia (or DBpedia) can be extended with the approach discussed in this paper. This holds, e.g., for YAGO and Wikidata. Moreover, the approach could also be applied to knowledge graphs created from other Wikis, such as DBkWik \cite{hertling2018dbkwik}, or used with different hierarchies, such as the Wikipedia Bitaxonomy \cite{flati2014two} or WebIsALOD \cite{hertling2017webisalod}. Hence, Cat2Ax has general potential which goes beyond DBpedia and Wikipedia.

\bibliographystyle{splncs04}
\bibliography{references}
\end{document}